# Density of States Extracted from Modified Recursion Relations

H. Bahlouli[a,+], A. D. Alhaidari[b], M. S. Abdelmonem[a]
[a] *Physics Department, King Fahd University of Petroleum & Minerals, Dhahran 31261, Saudi Arabia*
[b] *Shura Council, Riyadh 11212, Saudi Arabia*

We evaluate the density of states (DOS) associated with tridiagonal symmetric Hamiltonian matrices and study the effect of perturbation on one of its entries. Analysis is carried out by studying the resulting three-term recursion relation and the corresponding orthogonal polynomials of the first and second kind. We found closed form expressions for the new DOS in terms of the original one when perturbation affects a single diagonal or off-diagonal site or a combination of both. The projected DOS is also calculated numerically and its relation to the average DOS is explored both analytically and numerically.

## I. INTRODUCTION

The density of states (DOS) in condensed matter and statistical physics is a property that describes or quantifies how closely packed energy levels are in a given physical system. It is often expressed as a function of the energy $E$ or the wave vector $|\vec{k}|$ and denoted by $\rho(E)$. The energy $E$ is related to $|\vec{k}|$ through a dispersion relation, which depends on the details of the system at hand. The quantity $\rho(E)dE$ represents the number of allowed energy levels, within the energy range $E$ to $E+dE$. Mathematically the discrete DOS projected on state $\left|j\right\rangle$ is defined by

$$\rho_j(E) = \sum_m \left|\left\langle m \middle| j \right\rangle\right|^2 \delta(E - E_m), \qquad (1)$$

where the sum runs over all discrete spectrum of the Hamiltonian with eigenvalue $E_m$ and corresponding eigenstates $\left|m\right\rangle$. The projected DOS describes how a particular orbital $\left|j\right\rangle$ couples to the entire system. Physically, the local DOS on a given orbital $\left|j\right\rangle$ describes the intensity of each eigenstate of the system on this particular orbital. This quantity is of physical interest and can be measured by means of tunneling probes [1]. This projected DOS is related to the imaginary part of the Green function by

$$\rho_j(E) = -\frac{1}{\pi}\lim_{\varepsilon\to 0}\left[\operatorname{Im}\left\langle j\right|\tfrac{1}{E+i\varepsilon-H}\left|j\right\rangle\right] = \frac{1}{\pi}\lim_{\varepsilon\to 0}\left[\operatorname{Im} G_{jj}(E+i\varepsilon)\right], \qquad (2)$$

where $G$ is the Green function of the system. An advantage of the Green function approach is that it provides a direct method for calculating the DOS as evidenced by (2). In addition the Green function technique, which constitutes the analytic basis of the present work, is mathematically elegant as can be seen from the results obtained in this work.

+ Corresponding Author: email: bahlouli@kfupm.edu.sa.

The method used in this work to extract the DOS relies on the fact that we deal with Hamiltonian representations that are of tridiagonal nature. However this restriction does not limit the validity of our results since the matrix representations of any Hamiltonian operators can always be brought to a tridiagonal form in certain basis using the Lanczos method [2]. Actually the importance of tridiagonal Hamiltonians is due to the fact that by Lanczos method it is possible to map any quantum mechanical problem into a semi-infinite 1D chain represented by a tridiagonal Hamiltonian. This is the reason why tridiagonal Hamiltonians have special relevance in quantum mechanics [3]. Thus, starting with a seed unit vector $|\phi_0\rangle$, the Lanczos method [2] generates a set of orthonormal basis $\{|\phi_n\rangle\}_{n=0}^{\infty}$ that tridiagonalizes a given Hamiltonian

$$\langle\phi_n|H|\phi_m\rangle = b_{n-1}\delta_{n,m+1} + a_n\delta_{n,m} + b_n\delta_{n,m-1}. \tag{3}$$

For irreducible hermitian representations of the Hamiltonians, all $a_n$ are real and $b_n \neq 0$ for all $n$. Moreover, $b_n$ can be brought into real form by just including the appropriate phases in the corresponding basis functions. Expanding the wavefunction in this basis as $|\psi(E)\rangle$ = $\sum_n p_n(E)|\phi_n\rangle$ and inserting this expression in the stationary Schrödinger equation, $H|\psi\rangle$ = $E|\psi\rangle$, one obtains the following three term recursion relation for the expansion coefficients $\{p_n(E)\}$

$$E\,p_n(E) = b_{n-1}p_{n-1}(E) + a_n p_n(E) + b_n p_{n+1}(E) \quad ; \quad n \geq 1 \tag{4}$$

The solution of this recursion, $\{p_n(E)\}$, are defined modulo an arbitrary function of $E$. The un-normalized wave function is defined by fixing this arbitrariness with the standard normalization, $p_0(E) = 1$. This choice results in $p_n(E)$ being a polynomial of degree $n$ in $E$. Consequently, the expansion coefficients $\{p_n(E)\}$ form a set of polynomials orthogonal with respect to the projected density of states $\rho_0(E)$

$$\int_a^b \rho_0(E)p_n(E)p_m(E)dE = \delta_{mn}, \tag{5}$$

where the spectrum is assumed to be confined to the energy band $E \in [a,b]$. The norm of the wave function then becomes

$$|\psi(E)|^2 = \sum_{n=1}^{N} |p_n(E)|^2. \tag{6}$$

This norm can be looked at as being the signature of the nature of the state at energy $E$. Strongly localized states have normalization that is independent of the size $N$ of the system while weakly localized and extended states have normalizations which scale with the size of the system as $N^{\alpha}$ with $\alpha$ positive. Thus the normalization constant of the wave function $\psi(E)$ can be used as a criterion for distinguishing between localized and delocalized states [4]. Equation (2) gives $\rho_0(E)$ as follows

$$\rho_0(E) = \lim_{\varepsilon\to 0^+} \frac{1}{2\pi i}\left[G_{00}(E+i\varepsilon) - G_{00}(E-i\varepsilon)\right]. \tag{7}$$

That is, $\rho_0(E)$ is related to the discontinuity across the cut of the element $G_{00}(z)$ = $\langle\phi_0|G(z)|\phi_0\rangle$ on the segment of the real line $E \in [a,b]$ in the complex energy plane. The Green's function is formally defined in the complex $z$-plane as $G(z) = (H - zI)^{-1}$, where $I$ is the unit matrix. The task of obtaining $G(z)$ by inverting the infinite matrix $(H - zI)$ is often difficult. Furthermore, in realistic cases it is possible to obtain only a finite number

of elements $\{H_{nm}\}_{n,m=0}^{N-1}$. The infinite Hamiltonian matrix is then reduced to a finite $N\times N$ matrix $\tilde{H}$ which possesses a finite number of eigenvalues and normalized eigenvectors. If $\tilde{I}$ is $N\times N$ unit matrix, then $\tilde{G}(z)=(\tilde{H}-z\tilde{I})^{-1}$ is also an $N\times N$ matrix. However, $\tilde{G}_{00}(z)$ lacks the analytic cut structure necessary to define a density through relation (7). Instead $\tilde{G}_{00}(x+i0)$ possesses a set of poles and zeros which tries to mimic the cut structure of $G_{00}(z)$ in the complex energy plane.

In a recent paper by the co-Authors [5], three approximation methods were introduced to extract highly accurate density information (over a continuous range of energy) from $\tilde{G}$ without a need for knowledge of the asymptotic values of the coefficients $a_n$ and $b_n$. The first method, which is called the *Analytic Continuation* method [6], is based on the fact that $\tilde{G}(z)$ is a good approximation to $G(z)$ in a region of the complex energy plane away from the real axis. Thus the evaluation of $\tilde{G}(z)$ at a set of complex energies $\{z_i\}$ followed by analytic continuation to the real energy axis gives complex values from which the density can be extracted via relation (7). The second method, which is called the *Dispersion Correction* method [7], is based on the ability to define the analytic sense of discretization of the energy continuum and incorporate this information into the calculation. In this way it was possible to obtain density information over a continuous range of energies. The third method is the *Stieltjes Imaging* method [8] which is based on the fact that the density is related to the distribution of the eigenvalues and eigenvectors of the finite Hamiltonian $\tilde{H}$.

In this work, our aim is to study the effect of a particular kind of deformation of the tridiagonal Hamiltonian $H$. We want to investigate what happens to the projected density under the following modifications in the entries of $H$ [the recursion coefficients of relation (4)]

$$a_k \to a_k+\mu\text{, or } b_k \to \gamma b_k, \tag{8}$$

for a given non-negative integer $k$ and where $\mu$ and $\gamma$ are real and $\gamma\neq 0$. These two situations correspond to adding an impurity to our system, $\mu$ represents the strength of the scattering potential while $\gamma$ represents the scaling of the overlap integral between near neighbors. Real situations give rise to both effects at the same time since the inclusion of an impurity affects both the scattering potential and the overlap between near neighbor atoms.

The paper is organized as follows: In section II, we discuss the projected DOS and its numerical computation. In section III, we introduce the one-term perturbation (8) and study its effect on the density of state function. The combined effect of both types of diagonal and off diagonal deformations is dealt with in a separate appendix. In section IV, we conclude by discussing our main results and possible extension of this work

## II. PROJECTED DENSITY OF STATES FOR FINITE SYSTEMS

As stated above, the matrix representation of $H$ in the basis $\{\phi_n\}_{n=0}^{\infty}$ is tridiagonal. That is, Eq. (3) gives

$$H=\begin{pmatrix} a_0 & b_0 & & & & & \\ b_0 & a_1 & b_1 & & \mathbf{0} & & \\ & b_1 & a_2 & b_2 & & & \\ & & b_2 & a_3 & b_3 & & \\ & & & \times & \times & \times & \\ & \mathbf{0} & & & \times & \times & \times \\ & & & & & \times & \times \end{pmatrix} \tag{9}$$

where $\{a_n, b_n\}$ are real and $b_n \neq 0 \ \ \forall n$. Therefore, analysis of the system could, equivalently, be carried out in the vector space spanned by the complete set of $L^2[a,b]$ functions $\{p_n(x)\}$. That is, we study the system in the energy space that carries the spectral decomposition (Fourier expansion) of the state function, $|\psi(E)\rangle = \sum_n p_n(E)|\phi_n\rangle$. For numerical computations, however, this space is truncated into a finite $N$-dimensional sub-space spanned by $\{p_n(x)\}_{n=0}^{N-1}$. The tridiagonal matrix (9) becomes a finite $N\times N$ matrix $H$. The real eigenvalues of $H$, designated as the set $\{\varepsilon_n\}_{n=0}^{N-1}$, are the zeros of the polynomial $p_N(x)$. An approximation of the density function, which we will denote by $\rho'(x)$, can be obtained as follows. Let's define the finite Green's function $g_{00}^{(N)}(z) \equiv \left[(H-z)^{-1}\right]_{00}$. It has the following continued fraction representation [3]

$$g_{00}^{(N)}(z)=\cfrac{-1}{z-a_0-\cfrac{b_0^2}{z-a_1-\cfrac{b_1^2}{z-a_2-\dots\cfrac{b_{N-2}^2}{z-a_{N-1}}}}} \tag{10}$$

In the limit as $N\to\infty$, $g_{00}^{(\infty)}(z)\equiv G_{00}(z)$ is an analytic function in the complex variable $z$ with a branch cut on the real line. However, in most realistic cases the exact expressions of $G_{00}(z)$ are not known. In such a case, one may use (10) (or any one of the many alternative expressions [3,9]) for the finite Green's function $g_{00}^{(N)}(z)$ in a numerical scheme to obtain an approximation for $G_{00}(z)$. Let's call this approximate function $G'_{00}(z)$. Then, we can write

$$\rho'_0(x)=\frac{1}{2\pi i}\mathrm{Im}[G'_{00}(x+i0)-G'_{00}(x-i0)]=\frac{1}{\pi}\mathrm{Im}\,G'_{00}(x+i0)\,. \tag{11}$$

The approximation may be carried out using the "Analytic Continuation" method [6]. However, there are other schemes to obtain the approximate density function. Some of these schemes are outlined in reference [5]. In all approximation schemes, one utilizes knowledge of the given $2N-1$ recursion coefficients $\{a_n\}_{n=0}^{N-1}$ and $\{b_m\}_{m=0}^{N-2}$ or equivalently the set of eigenvalues of the $N\times N$ tridiagonal matrix $H$ and one of its submatrices (the one obtained by deleting the first row and first column).

The above formulation can be generalized using the finite Green's function $g_{n,n}^{(N)}(z)$ which can be related to the $g_{00}^{(N)}(z)$ as follows

$$g_{n,n}^{(N)}(z) = [p_n(z)]^2 \left[ g_{00}^{(N)}(z) - g_{00}^{(n)}(z) \right], \tag{12}$$

where the squared term after the equal sign is identical to unity for $n = 0$. This Green's function is associated with the $N \times N$ matrix $H$ and one of its submatrices, which is obtained by deleting the $n^{\text{th}}$ row and $n^{\text{th}}$ column. The Green's function $G_{n,n}(z)$ is defined as the limit of $g_{n,n}^{(N)}(z)$ as $N \to \infty$. Therefore, relation (12) holds true also for $G_{n,n}(z)$ and $G_{00}(z)$, respectively. Associated with this Green's function is the following "generalized density":

$$\rho_n(x) = \frac{1}{\pi} \operatorname{Im} G_{nn}(x + i0)\,. \tag{13}$$

This expression can be interpreted as the local DOS at the $n$-th site of the 1D chain. In case the exact $G_{n,n}(z)$ is not known, one may use the above expressions (12) for the finite Green's function $g_{n,n}^{(N)}(z)$ in a numerical scheme to obtain an approximation for $G_{n,n}(z)$. The three schemes in reference [5] to calculate the density function could also be extended to obtain an approximation for the set of generalized density functions $\left\{\rho_n(x)\right\}_{n=0}^{N-1}$. $n = 0$ is the case discussed in the previous paragraph. Figure 1 shows some of these generalized densities for the model defined by the following recursion coefficients which are associated with a renormalized version of the Gegenbauer polynomials[+]

$$a_n = 0 \quad , \quad b_n = \frac{1}{2}\sqrt{\frac{(n+1)(n+2\lambda)}{(n+\lambda)(n+\lambda+1)}} \quad ; n = 0,1,\ldots \tag{14}$$

where $\lambda$ is a real parameter greater than 1/2 . In the graphs, $N = 20$ and $\lambda = 3/2$. The generalized densities were obtained using the "Stieltjes Imaging" method (crosses) [8] and compared to the exact densities (solid line). For this model the exact density $\rho_0(x)$ is known:

$$\rho_0(x) = \frac{1}{\sqrt{\pi}} \frac{\Gamma(\lambda+1)}{\Gamma(\lambda+1/2)} \left(1 - x^2\right)^{\lambda - 1/2} \qquad ; x \in [-1,+1] \tag{15}$$

As a simple illustration, we consider the case $\lambda = 1$ then $a_n = 0$ and $b_n = 1/2$ which results in the Chebyshev polynomials of the second kind, $U_n(x)$, and the associated DOS is given by

$$\rho_0(x) = \frac{2}{\pi}\sqrt{1 - x^2}\,. \tag{16}$$

Taking the limit as $N \to \infty$ in (12), then the imaginary part will relate the projected DOS to $\rho_0(x)$ as follows

$$\rho_n(x) = \frac{1}{\pi} \operatorname{Im} G_{n,n}(x + i0) = [p_n(x)]^2 \rho_0(x)\,. \tag{17}$$

In the derivation of this result we used the fact that $g_{00}^{(n)}(z)$ has a finite number of simple poles with no branch cuts and hence has zero imaginary part. The averaged DOS, which gives equal weight to each eigenvalue, is defined by

$$\rho(x) = \frac{1}{N}\sum_{j=1}^{N} \rho_j(x) = \rho_0(x) \frac{1}{N}\sum_{j=1}^{N} \left[p_j(x)\right]^2\,. \tag{18}$$

---

[+] These are written in terms of the Gegenbauer polynomials, $C_n^{\lambda}(x)$, as $p_n(x) = \sqrt{\frac{\Gamma(2\lambda)}{\lambda}}\sqrt{\frac{(n+\lambda)\Gamma(n+1)}{\Gamma(n+2\lambda)}}\, C_n^{\lambda}(x)$.

In the limit $N \to \infty$ and using the fact that $p_n(x) = U_n(x)$ are just the Chebyshev polynomials of the second kind which satisfy the following sum rule

$$\lim_{n\to\infty} \frac{1}{n} \sum_{j=1}^{n} U_j^2(x) = \frac{1}{2\left(1-x^2\right)} . \tag{19}$$

Then (18) becomes

$$\rho(x) = \rho_0(x) \lim_{N\to\infty} \frac{1}{N} \sum_{j=1}^{N} \left[U_n(x)\right]^2 = \frac{1}{\pi} \frac{1}{\sqrt{1-x^2}} \qquad ; x \in [-1,+1] \tag{20}$$

This relation is also verified numerically in Fig. 2 where we show the sum of all projected DOS for $N$ = 5, 10, 25 (colored solid lines) and the averaged DOS as defined by Eq. (20) (dashed black line).

## III. DENSITY OF STATES ASSOCIATED WITH MODIFIED RECURSION RELATIONS

In this section we would like to study the effect of changing a single element in the tridiagonal Hamiltonian matrix $H$ which amounts to a change in one coefficient of the recursion relation. This modification can take place with the diagonal element of $H$ which physically represents a modification in the scattering potential and consequently can represent the effect of a single impurity scattering on the spectral density of the original host lattice. We can also modify the off diagonal element of the matrix $H$ which amounts physically to changing the overlap integral between neighboring sites. The net effect of these perturbations will certainly depend on the location of the modified site along the 1D chain. Thus our study is concerned with an arbitrary single parameter modification of the recursion coefficients. The real impurity situation will involves changes in both the diagonal and off diagonal elements, such a situation is just a linear combination of the two separate cases due to the linear nature of the eigenvalue problem which is validated by the superposition principle. Details about this general case can be found in the appendix.

### A. Perturbation of the *k*-th diagonal element of $H(a_k)$

We consider a 1-D chain represented by the tridiagonal Hamiltonian $H$ or, equivalently, by the general three-term recursion relation

$$\left(a_n - x\right) S_n\left(x\right) + b_{n-1} S_{n-1}\left(x\right) + b_n S_{n+1}\left(x\right) = 0 . \tag{21}$$

We would like to study the effect of changing arbitrarily a single site potential which is equivalent to changing

$$a_k \to \hat{a}_k = a_k + \mu , \qquad k = 0,1,2\ldots \tag{22}$$

This situation corresponds to adding a potential scattering of strength $\mu$ at the $k$-th site of the 1D chain. The solution of (21) is known (modulo an arbitrary non-zero function of $x$) to be polynomials of the “first kind” $p_n(x)$ having the initial conditions

$$p_0 = 1 \quad ; \quad p_1 = \frac{x - a_0}{b_0} . \tag{23}$$

Nonetheless, there exists another independent solution in terms of polynomials of the “second kind”, referred to by $q_n(x)$, that obey the same recursion relation (21) but with a different initial conditions

$$q_0 = 0 \quad ; \quad q_1 = \frac{1}{b_0} . \tag{24}$$

That is, $q_n(x)$ is, in fact, not a solution of (21) but of the following inhomogeneous recursion

$$(a_n - x) q_n(x) + b_{n-1} q_{n-1}(x) + b_n q_{n+1}(x) = \delta_{n0} . \tag{25}$$

It is clear that the recursion relation is not affected by the change (22) for $n \le k$ so that our new polynomials called $\widehat{p}_n$ and $\widehat{q}_n$ satisfy

$$\widehat{p}_n = p_n ; \widehat{q}_n = q_n \text{ for } n \le k . \tag{26}$$

The new polynomials are calculated as

$$\widehat{S}_{n+1}(x) = \left( \frac{x - \widehat{a}_n}{b_n} \right) \widehat{S}_n(x) - \frac{b_{n-1}}{b_n} \widehat{S}_{n-1}(x) , \tag{27}$$

with $\widehat{S}_n = \widehat{p}_n , \widehat{q}_n$ and $\widehat{S}_n = S_n$ for $n \le k$. Analysis of this recursive relation for $n > k$ gives the following result

$$\widehat{S}_{k+n}(x) = S_{k+n}(x) - \frac{\mu}{b_k} p_{n-1}^{(k+1)}(x) S_k(x); \; n \ge 1 , \tag{28}$$

where $\left\{ p_n^{(m)}, q_n^{(m)} \right\}$ are the polynomials associated with the tridiagonal matrix obtained from the original one, *H*, by deleting the first *m* rows and *m* columns ( also called the *m*-th “abbreviated” tridiagonal matrix [10]), and $p_n = p_n^{(0)}$ and $q_n = q_n^{(0)}$. That is, $p_n^{(m)}$ and $q_n^{(m)}$ are polynomials of the first- and second-kind satisfying the following three-term recursion relation

$$(a_{m+n} - x) S_n(x) + b_{m+n-1} S_{n-1}(x) + b_{m+n} S_{n+1}(x) = 0 , \qquad n \ge 1 \tag{29}$$

with initial conditions given by

$$p_0^{(m)} = 1 ; \; p_1^{(m)} = \frac{x - a_m}{b_m} \text{ and } q_0^{(m)} = 0 ; \; q_1^{(m)} = \frac{1}{b_m} . \tag{30}$$

It is worth mentioning that (28) is valid for $\widehat{S}_n = \widehat{p}_n$ or $\widehat{q}_n$ while (29) is valid for $S_n = p_n^{(m)}$ or $q_n^{(m)}$ .

In order to find a closed form relationship between the new DOS and the unperturbed one, we need to establish a relationship between the new and original Green functions. One way of calculating the Green’s function $G_{00}(z)$ is by taking the limit as $N \to \infty$ of Eq. (10):

$$G_{00}(z) = \cfrac{-1}{z - a_0 - \cfrac{b_0^2}{z - a_1 - \cfrac{b_1^2}{z - a_2 - ....}}} \tag{31}$$

Now, given a non-negative integer *k* we consider the following deformation

$$a_k \to a_k + \mu , \tag{32}$$

where $\mu$ is a real parameter. This deformation changes the two recursion relations above resulting in the new set of orthogonal polynomials given by relation (28). Now, we want to relate the deformed Green’s function $\widehat{G}^k_{00}(z)$ to the original one $G_{00}(z)$. Using Eq. (31), we can write for the new Green function

$$\hat{G}^{k}_{00}(z) = \frac{-1}{z - a_0 - \cfrac{b_0^2}{z - a_1 - \ldots - \cfrac{b_{k-1}^2}{z - a_k - \mu + b_k^2 T_{k+1}(z)}}} \tag{33}$$

where $T_m(z)$ is the "terminator", which is defined by the infinite continued fraction

$$T_m(z) = \frac{-1}{z - a_m - \cfrac{b_m^2}{z - a_{m+1} - \cfrac{b_{m+1}^2}{z - a_{m+2} - \ldots.}}} \tag{34}$$

However, we can also rewrite the original Green's function in Eq. (31) as

$$G_{00}(z) = \frac{-1}{z - a_0 - \cfrac{b_0^2}{z - a_1 - \ldots - \cfrac{b_{k-1}^2}{z - a_k + b_k^2 T_{k+1}(z)}}} \tag{35}$$

Solving for $T_{k+1}(z)$ from this equation one obtains

$$z - a_k + b_k^2 T_{k+1}(z) = \frac{b_{k-1}^2}{z - a_{k-1} - \cfrac{b_{k-2}^2}{z - a_{k-2} - \ldots. - \cfrac{b_0^2}{z - a_0 + \cfrac{1}{G_{00}(z)}}}} \tag{36}$$

substituting this result in Eq. (33) we obtain the sought after relation

$$\hat{G}^{k}_{00}(z) = \frac{-1}{z - a_0 - \cfrac{b_0^2}{z - a_1 - \ldots - \cfrac{b_{k-2}^2}{z - a_{k-1} + \cfrac{1}{\cfrac{\mu}{b_{k-1}^2} - \cfrac{1}{z - a_{k-1} - \cfrac{b_{k-2}^2}{z - a_{k-2} - \ldots - \cfrac{b_0^2}{z - a_0 + \cfrac{1}{G_{00}(z)}}}}}}}} \tag{37}$$

This expression, although very well suited for stable numerical computations, could be simplified analytically. To this end, we use relations of the continued fraction to its terminator as ratios involving polynomials of the first and second kind. After few iterative manipulations, one can show that

$$z - a_{k-1} - \frac{b_{k-2}^2}{z - a_{k-2} - \ldots. - \cfrac{b_0^2}{z - a_0 + T(z)}} = b_{k-1} \frac{q_k(z)T(z) + p_k(z)}{q_{k-1}(z)T(z) + p_{k-1}(z)} \tag{38}$$

$$z - a_1 - \frac{b_1^2}{z - a_2 - \ldots - \cfrac{b_{k-1}^2}{z - a_k + T(z)}} = b_0 \frac{Q_k^{(k+1)}(z)T(z) + P_k^{(k+1)}(z)}{Q_{k-1}^{(k+1)}(z)T(z) + P_{k-1}^{(k+1)}(z)} \tag{39}$$

where $\left\{P_n^{(m)}, Q_n^{(m)}\right\}_{n=0}^{m-1}$ are the polynomials associated with the finite $m \times m$ tridiagonal matrix

$$H^{(m)} = \begin{pmatrix} a_{m-1} & b_{m-2} & & & & & \\ b_{m-2} & a_{m-2} & b_{m-3} & \mathbf{0} & & & \\ & b_{m-3} & a_{m-3} & \times & & & \\ & & \times & \times & \times & & \\ & & & \times & \times & b_1 & \\ & \mathbf{0} & & & b_1 & a_1 & b_0 \\ & & & & & b_0 & a_0 \end{pmatrix} \tag{40}$$

which is obtained from the original one, $H$ as defined by equation (9), by keeping the first $m$ rows and $m$ columns and performing the inversion $a_n \to a_{m-n-1}$ and $b_n \to b_{m-n-2}$. That is, $P_n^{(m)}$ and $Q_n^{(m)}$ are polynomials of the first- and second-kind satisfying the following three-term recursion relation

$$\left(a_{m-n-1} - x\right) S_n(x) + b_{m-n-1} S_{n-1}(x) + b_{m-n-2} S_{n+1}(x) = 0 \,, \quad n = 1, 2, .., m-2 \,, \tag{41}$$

with initial values given by

$$P_0^{(m)} = 1 \; ; \; P_1^{(m)} = \frac{x - a_{m-1}}{b_{m-2}} \text{ and } Q_0^{(m)} = 0 \; ; \; Q_1^{(m)} = \frac{1}{b_{m-2}} \,. \tag{42}$$

Inserting the results from Eqs. (38) and (39) into Eq. (37), we can rewrite the modified Green function in terms of the original one as follows

$$\begin{aligned} \widehat{G}^k_{00}(z) &= -\frac{\widehat{A}_k(z) G_{00}(z) + \widehat{B}_k(z)}{\widehat{C}_k(z) G_{00}(z) + \widehat{D}_k(z)} \\ \widehat{A}_k &= Q_k^{(k+1)} \left(\mu p_k + b_k p_{k+1}\right) - P_k^{(k+1)} p_k \\ \widehat{B}_k &= Q_k^{(k+1)} \left(\mu q_k + b_k q_{k+1}\right) - P_k^{(k+1)} q_k \\ \widehat{C}_k &= Q_{k+1}^{(k+1)} \left(\mu p_k + b_k p_{k+1}\right) - P_{k+1}^{(k+1)} p_k \\ \widehat{D}_k &= Q_{k+1}^{(k+1)} \left(\mu q_k + b_k q_{k+1}\right) - P_{k+1}^{(k+1)} q_k \end{aligned} \tag{43}$$

where $P_m^{(m)}$ and $Q_m^{(m)}$ are defined by $S_m(x) = \left(x - a_0\right) S_{m-1}(x) - b_0 S_{m-2}(x)$. That is, by taking $n = m - 1$ in Eq. (41) and defining $b_{-1} \equiv 1$ (for the special case $m$ = 1, $P_1^{(1)} = x - a_0$ and $Q_1^{(1)} = 1$). Thus, the new DOS and the old one are related by

$$\widehat{\rho}_k(E) = \frac{\widehat{B}_k(E)\widehat{C}_k(E) - \widehat{A}_k(E)\widehat{D}_k(E)}{\left|\widehat{C}_k(E) G_{00}(E) + \widehat{D}_k(E)\right|^2} \rho_0(E) \quad ; \quad k = 0, 1, 2, ... \tag{44}$$

After some algebra and use of the Wronskian-like relation [9]

$$q_n p_{n-1} - q_{n-1} p_n = \frac{1}{b_{n-1}} \,. \tag{45}$$

We can simplify (44) to read as follows

$$\begin{aligned} \widehat{\rho}_k(E) &= \frac{Q_{k+1}^{(k+1)} P_k^{(k+1)} - Q_k^{(k+1)} P_{k+1}^{(k+1)}}{\left|\widehat{C}_k(E) G_{00}(E) + \widehat{D}_k(E)\right|^2} \rho_0(E) \\ &= \frac{\rho_0(E)}{\left|\widehat{C}_k(E) G_{00}(E) + \widehat{D}_k(E)\right|^2} \; ; \; k = 0, 1, 2, ... \end{aligned} \tag{46}$$

due to the fact that $P_n^{(m)}$ and $Q_n^{(m)}$ satisfy a Wronskian-like relation similar to (45)

$$Q_n^{(m)}P_{n-1}^{(m)} - Q_{n-1}^{(m)}P_n^{(m)} = \frac{1}{b_{m-n-1}}, \tag{47}$$

where $n = 1, 2, .., m-1$. However, for $n = m$ the right-hand side is equal to one due to the definition of $P_m^{(m)}$ and $Q_m^{(m)}$ given above. In Fig. 3, we compare this closed form of the DOS as given by Eq. (46) (solid line) to the numerical approximation (crosses). We used the model defined by the recursion coefficients in (14) for $\lambda = 2$ and took $\mu = .05$ (a.u.) and $k = 1, 3$. It is worth mentioning that for $k = 0$ we obtain the particular result found by Alhaidari [11]. This situation is analogous to the effect of potential scattering studied by Kondo [12] where it was found that the main effect of impurity potential scattering on the properties of the system reduces to an effective DOS

$$\hat{\rho}_k(E) = \frac{\rho_0(E)}{\left|1 + \mu\, G_{00}(E)\right|^2}. \tag{48}$$

A result which could be deduced from the general formula (46) for $k = 0$.

### B. Perturbation of the *k*-th off diagonal term in *H*($b_k$)

Going back to the original recursion

$$\left(a_n - x\right)S_n\left(x\right) + b_{n-1}S_{n-1}\left(x\right) + b_nS_{n+1}\left(x\right) = 0, \tag{49}$$

we would like to study the effect of changing arbitrarily a single off-diagonal element of $H$ using a scaling parameter $\gamma$

$$b_k \to \tilde{b}_k = \gamma b_k, \quad k = 0,1,2,\ldots \tag{50}$$

where $\gamma \neq 0$. The new polynomials are then defined by

$$\tilde{S}_{n+1}\left(x\right) = \left(\frac{x - a_n}{\tilde{b}_n}\right)\tilde{S}_n\left(x\right) - \frac{\tilde{b}_{n-1}}{\tilde{b}_n}\tilde{S}_{n-1}\left(x\right), \tag{51}$$

with $\tilde{S}_n = \tilde{p}_n, \tilde{q}_n$ and $\tilde{S}_n = S_n$ for $n \leq k$. Now, after few iterations, we can generalize our previous result and write it in the following form

$$\tilde{S}_{k+n}\left(x\right) = \frac{1}{\gamma}S_{k+n}\left(x\right) + b_k\left(\frac{1}{\gamma} - \gamma\right)q_{n-1}^{(k+1)}(x)S_k\left(x\right) \quad ; \quad n \geq 1 \tag{52}$$

Similar to the previous case, we need to relate the new Green's function and density to the original ones. Performing a treatment similar to that which was carried out in subsection A above, we obtain the following continued faction representation of the Green's function associated with this deformed system

$$\tilde{G}_{00}^{k}(z) = \cfrac{-1}{z - a_0 - \cfrac{b_0^2}{z - a_1 - \ldots - \cfrac{b_{k-2}^2}{z - a_{k-1} - \cfrac{1}{\cfrac{1-\gamma^2}{b_{k-1}^2}\left(z - a_k\right) + \cfrac{\gamma^2}{z - a_{k-1} - \cfrac{b_{k-2}^2}{z - a_{k-2} - \ldots - \cfrac{b_0^2}{z - a_0 + \cfrac{1}{G_{00}(z)}}}}}}}} \tag{53}$$

Using the results obtained in Eqs. (38) and (39) we end up with the following equivalent expression

$$\tilde{G}_{00}^{k}(z) = -\frac{\tilde{A}_k(z)G_{00}(z) + \tilde{B}_k(z)}{\tilde{C}_k(z)G_{00}(z) + \tilde{D}_k(z)}$$

$$\begin{aligned}
\tilde{A}_k &= \gamma^2 b_k Q_k^{(k+1)} p_{k+1} - P_k^{(k+1)} p_k \\
\tilde{B}_k &= \gamma^2 b_k Q_k^{(k+1)} q_{k+1} - P_k^{(k+1)} q_k \\
\tilde{C}_k &= \gamma^2 b_k Q_{k+1}^{(k+1)} p_{k+1} - P_{k+1}^{(k+1)} p_k \\
\tilde{D}_k &= \gamma^2 b_k Q_{k+1}^{(k+1)} q_{k+1} - P_{k+1}^{(k+1)} q_k
\end{aligned} \tag{54}$$

Consequently, the new density becomes related to the original as

$$\tilde{\rho}_k(E) = \frac{\gamma^2 \rho_0(E)}{\left|\tilde{C}_k(E)G_{00}(E) + \tilde{D}_k(E)\right|^2} \quad ; k = 0,1,2,... \tag{55}$$

Figure 4 shows a comparison between this closed form (solid line) and the numerical computation of the DOS (crosses). The same model is used as that of Fig. 3 with $\gamma = 0.95$. The agreement between the closed form solution we obtained and the direct numerical computation of the modified DOS obtained using the "Stieltjes Imaging" method [8] is clear from this figure.

## IV. DISCUSSION OF RESULTS

A general structure resulted in the three situations at hand where all modified orthogonal polynomials can be derived from the old ones with an additional deformation that depends on the product of the $k$-th order polynomial and a new orthogonal polynomial. The new ones are associated with the $k$-th "abbreviated" tridiagonal matrix obtained from the original one by deleting the first $k$+1 columns and $k$+1 rows. The numerical implementation of the corresponding DOS was easily carried out for few cases while closed form formulae were derived which relate the deformed Green's functions and DOS to the original ones. The general structure of the modified Green functions, as expressed in formulae (43), (54) and (A1), reflects the power and elegance of the Green function approach in dealing with this type of single impurity problem. In addition this technique provided us with a direct access to the modified DOS which was expressed in terms of the unperturbed one. As mentioned elsewhere in the mathematics literature, perturbation on the diagonal terms ($a_n$) are called co-recursive polynomials while dilation of the off-diagonal ($b_n$) are called co-dilated polynomials [13]. We need to stress here that the construction of new orthogonal polynomials by changing and shifting recurrence coefficients has been a subject of great interest in the mathematics literature as evidenced in [13] and references therein. The mathematical question usually addressed is concerned with the effect of changing the recursion coefficients on the orthogonal polynomials [14]. It is our hope that the above derivations are more transparent and useful to the physics community. We particularly stressed the effect of such deformations on the DOS of the system, a quantity to which many physical properties are related. The single site deformation we dealt with can mimic the effect of inserting an impurity (defect) in a given 1D host lattice represented by a tight binding model. It is the hope that our results will make significant contributions to studying the effect of such scattering potential provided by the impurity as a function of its location along the 1D chain. In particular, the effects of this impurity on the band structure of the host lattice and on its potential contribution to surface bound states. A similar study for 1D system with band gaps will be very beneficial in probing impurity effects on band edges and bound states within the gap region.

## ACKNOWLEDGMENTS

The authors acknowledge the support of King Fahd University of Petroleum and Minerals under project FT-2005/11.

## APPENDIX: Perturbation of the *k*-th diagonal and off-diagonal terms in $H(a_k,b_k)$

If we now consider the simultaneous deformations $a_k \to a_k + \mu$ and $b_k \to \gamma b_k$, then using the same procedure in Sec. III we obtain, after a lengthy algebra, the following general results

$$\begin{aligned}
\bar{G}_{00}^{k}(z) &= -\frac{\bar{A}_k(z)G_{00}(z)+\bar{B}_k(z)}{\bar{C}_k(z)G_{00}(z)+\bar{D}_k(z)} \\
\bar{A}_k &= Q_k^{(k+1)}\left(\mu p_k + \gamma^2 b_k p_{k+1}\right) - P_k^{(k+1)} p_k \\
\bar{B}_k &= Q_k^{(k+1)}\left(\mu q_k + \gamma^2 b_k q_{k+1}\right) - P_k^{(k+1)} q_k \\
\bar{C}_k &= Q_{k+1}^{(k+1)}\left(\mu p_k + \gamma^2 b_k p_{k+1}\right) - P_{k+1}^{(k+1)} p_k \\
\bar{D}_k &= Q_{k+1}^{(k+1)}\left(\mu q_k + \gamma^2 b_k q_{k+1}\right) - P_{k+1}^{(k+1)} q_k
\end{aligned} \tag{A1}$$

Additionally, the corresponding deformed density is related to the original one as follows

$$\bar{\rho}_k(E) = \frac{\gamma^2 \rho_0(E)}{\left|C_k(E)G_{00}(E)+D_k(E)\right|^2} \quad ; k = 0,1,2,... \tag{A2}$$

It is obvious that what we have obtained in subsections III-A and III-B above are special cases of these general results corresponding to $\gamma = 1$ and $\mu = 0$, respectively. Figure 5 shows two examples of the effect of this combined deformation on the DOS.

**FIGURES CAPTION:**

**Fig. 1:** The lowest four projected densities associated with the system defined by the recursion coefficients (14) with $\lambda$ = 3/2. These generalized densities were obtained using the "Stieltjes Imaging" method [8] (crosses) and compared to the exact densities (solid line) of Eq. (17). In the graphs, the approximation size is $N$ = 20.

**Fig. 2** (color online): The sum of all $N$ projected DOS associated with the model defined in (14) with $\lambda$ = 1 for $N$ = 5, 10, 25 (colored solid lines) and compared to the averaged DOS as given by Eq. (20) (dashed black line).

**Fig. 3:** The solid trace shows the deformed DOS for $a_k \to a_k + \mu$ as given by the analytic expression (46) and compared to the numerical approximation (crosses). We used the model defined by the recursion coefficients (14) with $\lambda$ = 2 and took $\mu$ = .05 (a.u.). Figure (3a) and (3b) corresponds to $k$ = 1 and $k$ = 3, respectively.

**Fig. 4** Same as Fig. 3 but for the scaling deformation $b_k \to \gamma b_k$ with $\gamma$ = .95.

**Fig. 5** The solid trace shows the generalized deformed DOS corresponding to $a_k \to a_k + \mu$ and $b_k \to \gamma b_k$ as given by the analytic expression (A2) and compared to the numerical approximation (crosses). We used the model defined by the recursion coefficients (14) with $\lambda$ = 3 and took $\mu$ = 0.1 and $\gamma$ = 0.9. Figure (5a) and (5b) corresponds to $k$ = 0 and $k$ = 1, respectively.

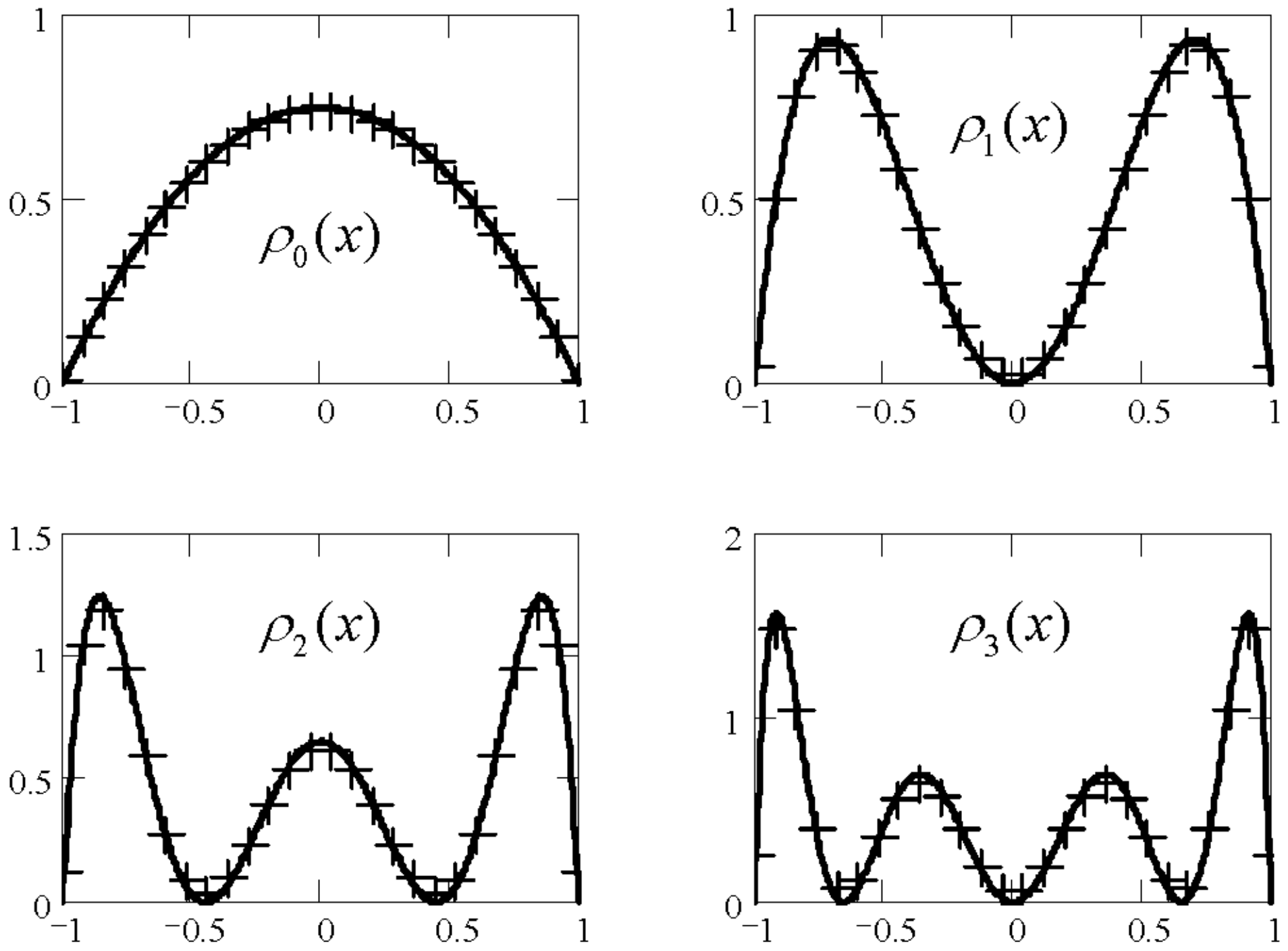

**Fig. 1**

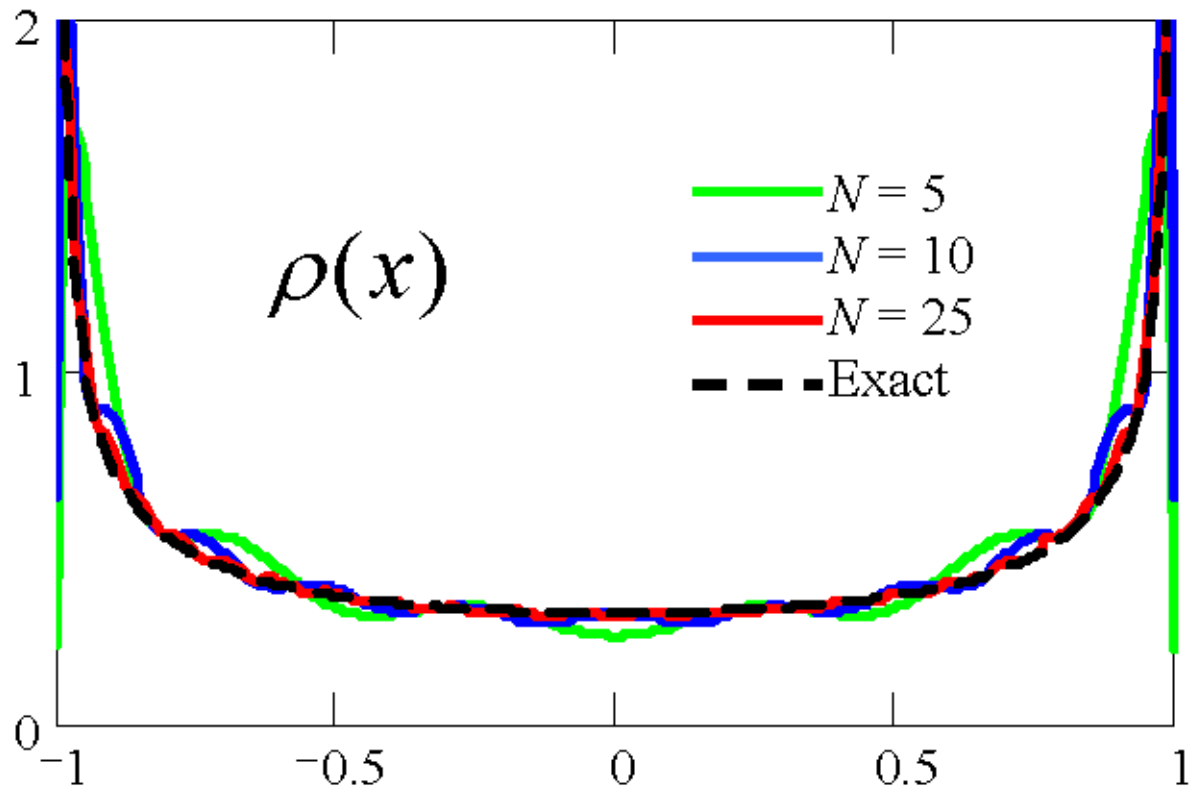

**Fig. 2**

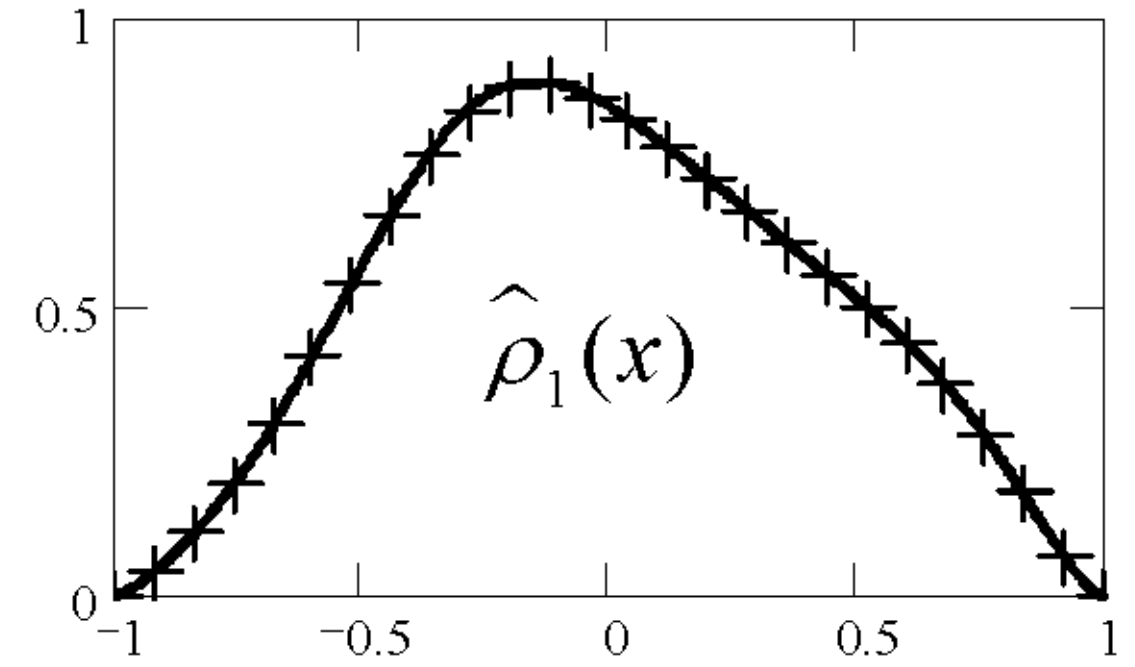

**Fig. 3a**

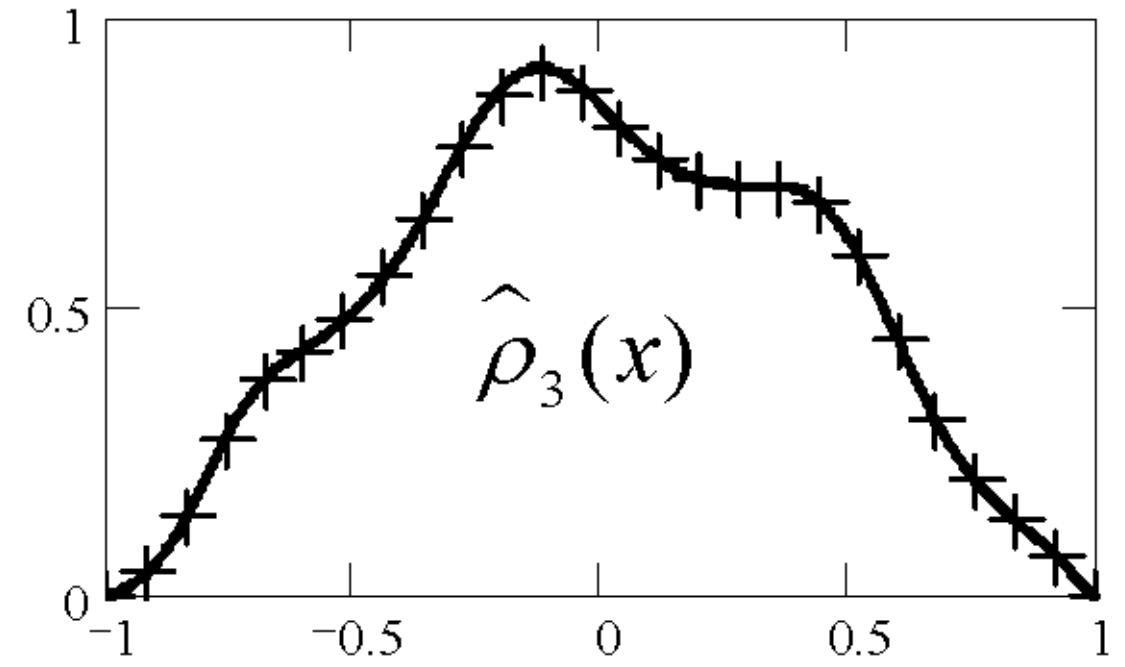

**Fig. 3b**

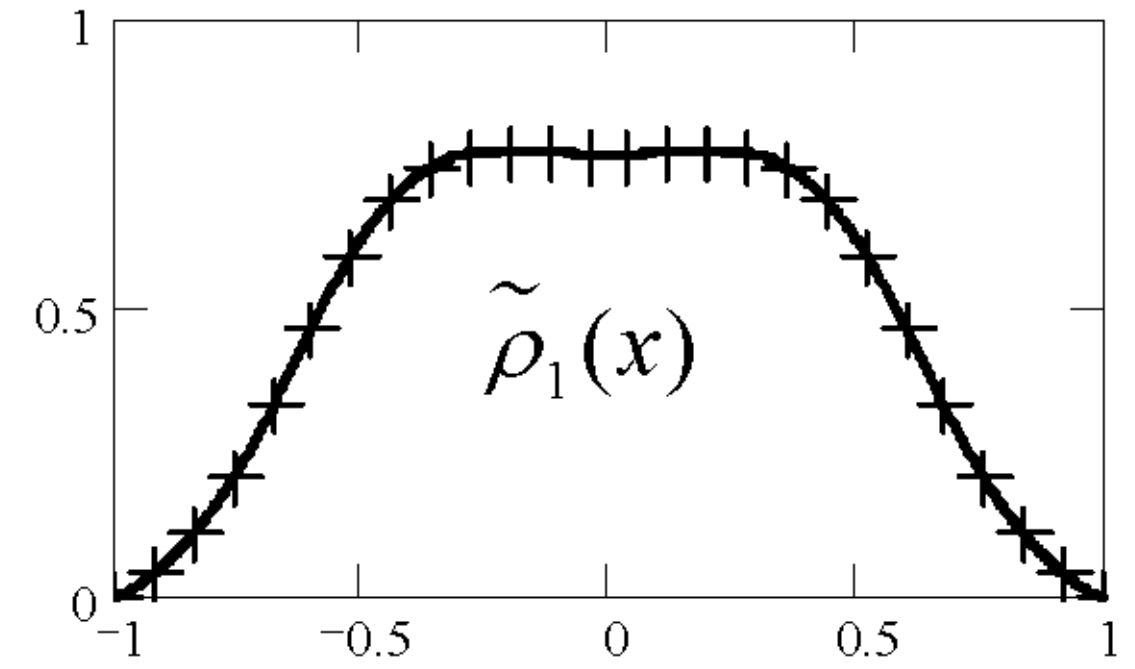

**Fig. 4a**

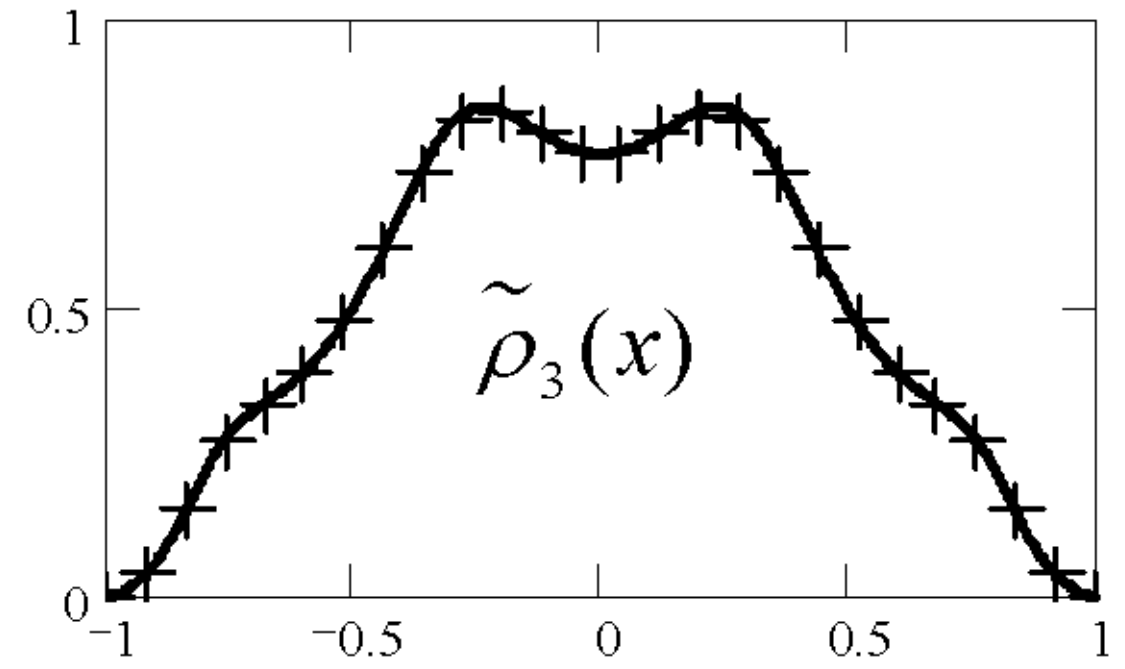

**Fig. 4b**

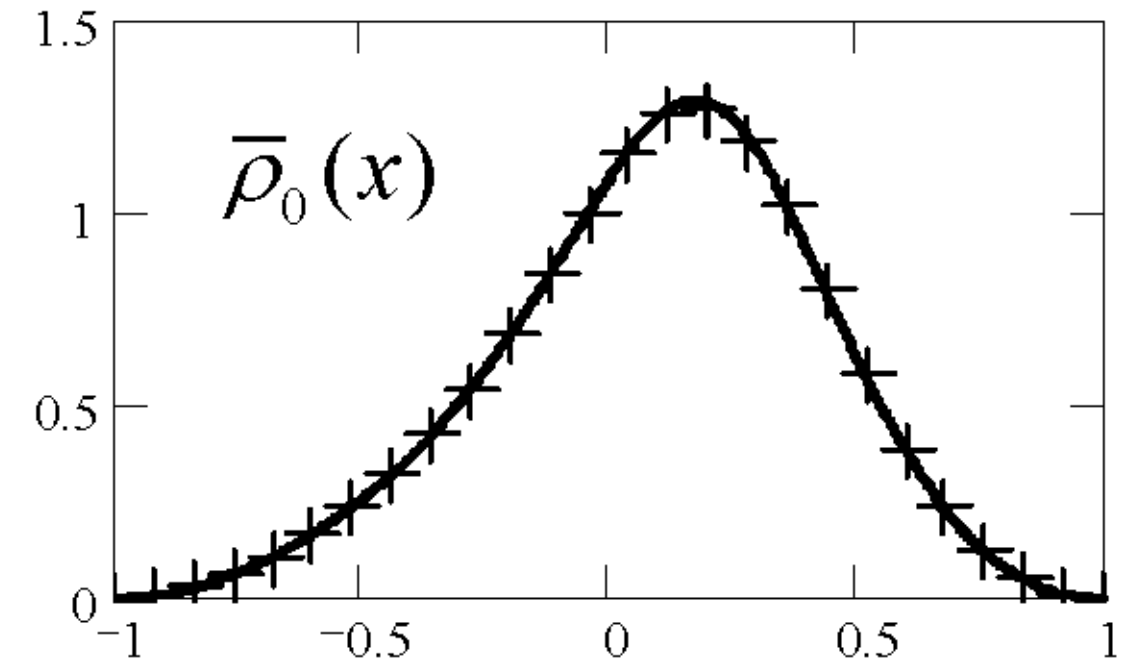

**Fig. 5a**

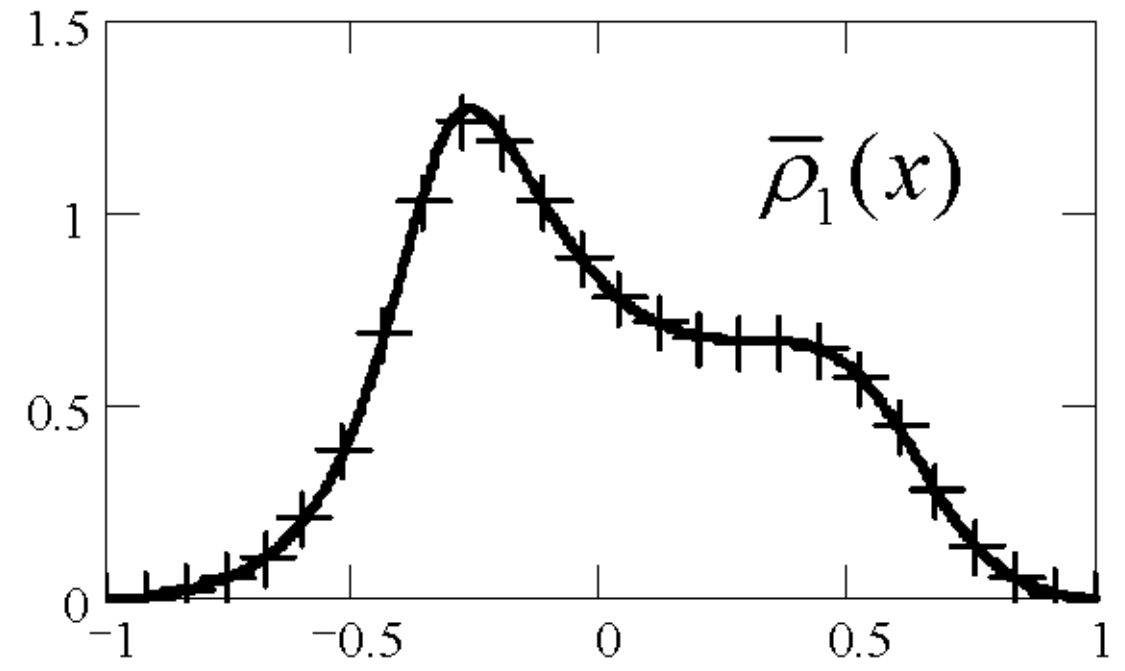

**Fig. 5b**